\documentclass[a4paper]{jpconf}
\usepackage{graphicx}
\newcommand {\beq}{\begin{equation}}
\newcommand {\eeq}{\end{equation}}
\newcommand {\bea}{\begin{eqnarray}}
\newcommand {\eea}{\end{eqnarray}}
\newcommand {\nn}{\nonumber \\}
\newcommand {\m}{\mu}
\newcommand {\n}{\nu}
\newcommand {\pl}{\partial}

\newcommand {\al}{\alpha}
\newcommand {\be}{\beta}

\newcommand {\la}{\lambda}
\newcommand {\La}{\Lambda}

\newcommand {\om}{\omega}

\newcommand {\ep}{\epsilon}
\newcommand {\vep}{\varepsilon}

\newcommand {\del}  {\delta}

\newcommand {\mn}{{\mu\nu}}

\newcommand {\half}{ {\frac{1}{2}} }

\newcommand {\Ecal}{{\cal E}}
\newcommand {\Fcal}{{\cal F}}
\newcommand {\Lcal}{{\cal L}}

\newcommand {\Pcal}{{\cal P}}
\newcommand {\Dcal}{{\cal D}}

\newcommand {\ptil} {{\tilde p}}
\newcommand {\ktil} {{\tilde k}}

\newcommand {\K}{{\bf K}}
\newcommand {\I}{{\bf I}}

\newcommand {\change} {\leftrightarrow}
\newcommand {\ra} {\rightarrow}

\newcommand {\pr}   {{\quad .}}
\newcommand {\com}  {{\quad ,}}
\newcommand {\q}    {\quad}

\newcommand {\PTP}  {{\it Prog.Theor.Phys.}}
\newcommand {\intfx} {{\int d^4x}}

\newcommand {\intpE} {{\int \frac{d^4p_E}{(2\pi)^4}}}
\newcommand {\intpL} {{\int_{\ptil\leq\Lambda} \frac{d^4p}{(2\pi)^4}}}

\newcommand {\Pla} {\frac{{\tilde p}}{\omega}}

\newcommand {\Tev} {\frac{{\tilde p}}{T}}

\begin{document}
\title{Casimir Energy of the Universe and the Dark Energy Problem}

\author{Shoichi Ichinose}

\address{
Laboratory of Physics, School of Food and Nutritional Sciences, 
University of Shizuoka\\
Yada 52-1, Shizuoka 422-8526, Japan
}

\ead{ichinose@u-shizuoka-ken.ac.jp}

\begin{abstract}
We regard the Casimir energy of the universe as the main contribution 
to the cosmological constant. Using 5 dimensional models of the 
universe, the flat model and the warped one, we calculate Casimir energy. 
Introducing the new regularization, called {\it sphere lattice regularization}, 
we solve the divergence problem. 
The regularization utilizes the closed-string configuration. 
We consider 4 different approaches: 
1) restriction of the integral region (Randall-Schwartz), 
2) method of 1) using the minimal area surfaces, 
3) introducing the weight function, 
4) {\it generalized path-integral}. 
We claim 
the 5 dimensional field theories are quantized properly and all divergences 
are renormalized. 
At present, it is explicitly demonstrated in the numerical 
way, not in the analytical way. The renormalization-group function ($\be$-function) 
is explicitly obtained. The renormalization-group flow of the cosmological 
constant is concretely obtained. 
\end{abstract}

\section{Introduction}
At present, the energy budget of the universe is not explained theoretically. 
We do not understand the dark matter and the dark energy. The latter one, which 
is the present main topic, 
is regarded as the same problem as the cosmological constant one, 
which is the long-lasting problem\cite{Wein89,Pad03}. 
As for the problem, Polyakov\cite{Pol82,Pol0709} made a notable conjecture 
that the cosmological constant, just like the QED coupling, flows to a small value 
in the IR region (screening phenomena). He made another comment\cite{Pol0912} that 
the dark energy, like the black body radiation 150 years ago, hides secrets of 
fundamental physics. These views about the problem look to be revived in the recent 
strong trend of the AdS/CFT (holography) approach to the condensed matter physics or 
to the viscous fluid dynamics(Navier-Stokes equation)\cite{BKLS1101,LysStr1104}. 

The difficulty of the problem is strongly related to the unsuccessful situation of the quantum gravity. 
It also has the long research history since Feynman initiated in 1963\cite{Fey63} until the present 
string or D-brane research. 
In the microscopic side, the gravitational interaction has, when quantized, 
serious UV-divergences\cite{tHooVel74}, while 
, in the macroscopic one, it has serious IR-divergences around the horizon 
(boundaries)\cite{Wit0106,Pol0709}.

About 2 years ago (2010 January), Verlinde\cite{Ver1001} made a shocking view about the role of the gravitational force. 
He made a very thoughtful analysis, including some Gedanken-experiments, 
about the gravitational force and 
finally reached the conclusion that the gravitational interaction (force) is {\it not} fundamental but 
is {\it emergent} (by some still-obscure mechanism). He referred some (not few) past literature about similar 
stand points such as Jacobson's\cite{Jac9504} and Padmanabhan's\cite{Pad0911}. The analysis demands very delicate treatment of the horizon which 
is the boundary of the theory. The physical quantities (IR-)diverge at the horizon. 
It strongly indicates the regularization problem in the near-horizon treatment where 
Hawking radiation (thermalization) occurs. 

The biggest discrepancy, in all physical quantities, between observation and theory 
appears in the cosmological constant $\la$. 
\bea
S=\int d^4x \sqrt{-g}\{ \frac{1}{16\pi G_N}(R-2\la) 
+\Lcal_{matter}\}\
\com\nn
\frac{\la_{th}}{\la_{obs}}\sim N_{DL}^{~2}\com\q  
N_{DL}\equiv M_{pl}R_{cos}\sim 6.\times 10^{60}
\com
\label{conc3}
\eea
where $M_{pl}$ ($\equiv 1/\sqrt{G_N}$) and $R_{cos}$ ($\equiv 1/H_0, H_0:$ Hubble constant) 
are Planck mass and the size of 
the universe respectively. $N_{DL}$ is an analog of Dirac's large number\cite{PD78}. 
\footnote{
The original Dirac's definition is [electromagnetic force]/[gravitational force]
=$\frac{\al\hbar c}{G_Nm_em_p}\sim 2.3\times 10^{39}~(\al=e^2/4\pi\vep_0\hbar c)$ 
where the force is that working 
between the electron and the proton in a H-atom. The substitute, in the present case, is naturally 
[size of the universe]/[Planck length]. 
Note that Dirac's value is approximately equal to $([\mbox{Planck mass}]/[\mbox{mass of nucleon}])^2$.
}
The cause of the theoretical difficulty lies in the two 
extreme-ends of mysterious branches of physics: the quantum gravity and 
the cosmology. 

Casimir energy is generally the vacuum energy of the {\it free}-part 
of the system dynamics. For the harmonic oscillator, it is the 
energy of the zero-point oscillation. By definition, it is 
independent of the coupling. It depends, however, on the {\it boundaries} 
and the system {\it topology}. It is caused by the quantum {\it fluctuation}. 
Because we have to deal with serious UV and IR divergences, highly-delicate 
{\it regularization} is required. 
The well-known one is that for the 4D electromagnetism (free wave theory). 
The system of two parallely-placed metalic plates separated by the length $l$, 
has Casimir energy as follows. 
\bea
\mbox{4D Space-time}:\ V_{Cas}(l)=\frac{B_4}{l^3},\ \ 
F_{Cas}(l)=-\frac{\pl V}{\pl l}=3\frac{B_4}{l^4},\ \ 
B_4(\mbox{4-th Bernoulli No})=-\frac{1}{30},\nn
\mbox{5D Space-time}:\ V_{Cas}(l)=-\frac{3}{32\pi^2}\frac{\zeta(5)}{l^4},\ \ 
F_{Cas}(l)=-\frac{\pl V}{\pl l}=-\frac{3}{8\pi^2}\frac{\zeta(5)}{l^5},\ \ 
\zeta(5)=1.03693\cdots
.
\label{ACpot}
\eea

\section{Casimir Energy of 5 Dimensional Electromagnetism}
For the electromagnetism in the flat 5D space-time 
($ds^2=\eta_\mn dx^\m dx^\n+dy^2,~\eta_\mn=\mbox{diag}(-1,1,1,1)$), 
Casimir energy is given by \cite{SI0801}
\bea
E_{Cas}(\La,l)=\intpL\int_0^ldy (F_f^-(\ptil,y)+4F_f^+(\ptil,y))\com\nn
F_f^\mp(\ptil,y)
=-\int_\ptil^\infty d\ktil\frac{\mp\cosh\ktil(2y-l)+\cosh\ktil l}{2\sinh(\ktil l)}
\com
\label{HK20}
\eea
where $\mp$ is the $y$-parity ($y\change -y$), $l$ is the periodicity 
$y\ra y+2l$, and $\La$ is the momentum cut-off. $\ptil$ is the 
magnitude of the 4D Euclidean momentum $(p_1,p_2,p_3,p_4=ip_0)$. 
This is consistent with Ref.\cite{AC83}. 

As for the warped (AdS$_5$) geometry, 
$ds^2=(\eta_\mn dx^\m dx^\n+dz^2)/\om^2z^2$ ($\om$: 5D curvature), 
the corresponding one is given by \cite{SI0812}
\bea
-E^{\La,\mp}_{Cas}(\om,T)
=\left.\intpE\right|_{\ptil\leq\La}\int_{1/\om}^{1/T}dz~F_w^\mp(\ptil,z), 
F_w^\mp(\ptil,z)= \frac{1}{(\om z)^3}\int_{\ptil^2}^\infty\{G_k^\mp (z,z)\}dk^2,\nn
G_p^\mp(z,z')=\mp\frac{\om^3}{2}z^2{z'}^2
\frac{\{\I_0(\Pla)\K_0(\ptil z)\mp\K_0(\Pla)\I_0(\ptil z)\}  
      \{\I_0(\Tev)\K_0(\ptil z')\mp\K_0(\Tev)\I_0(\ptil z')\}
     }{\I_0(\Tev)\K_0(\Pla)-\K_0(\Tev)\I_0(\Pla)},
\label{HKA11}
\eea
where $\I_0$ and $\K_0$ are 0-th modified Bessel functions. 
For simplicity, we take the bulk particle mass M in such a way that 
$M^2=-4\om^2<0$. $T$ is defined by $T=\om\exp({-l\om})$.

\section{Behavior of Casimir Energy }
\begin{figure}[h]
\begin{minipage}{18pc}
\includegraphics[width=18pc]{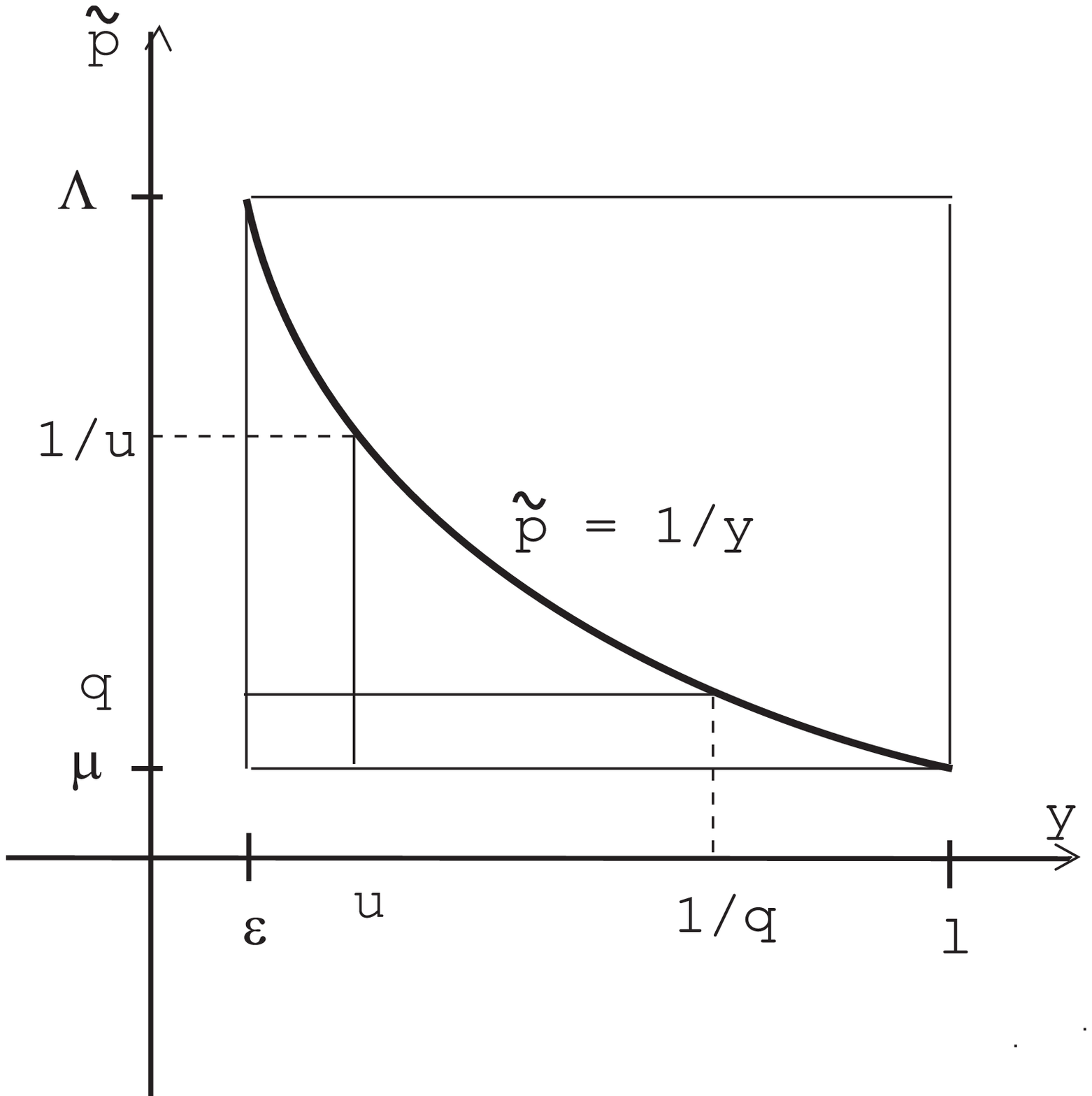}
\caption{\label{ypINTregion}
Space of (y,$\ptil$) for the integration (\ref{HK20}). The hyperbolic curve 
is used in (\ref{surfM1}). 
}
\end{minipage}\hspace{2pc}%
\begin{minipage}{18pc}
\includegraphics[width=18pc]{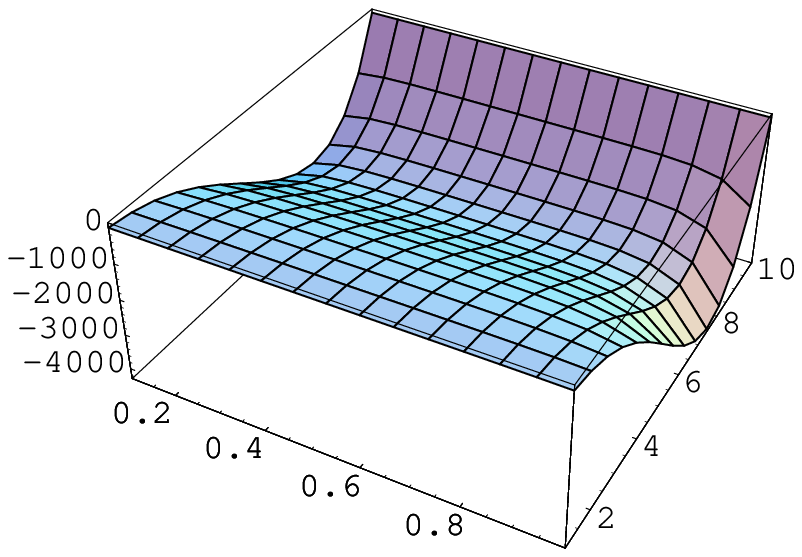}
\caption{\label{p3F10La}
Behaviour of $\ptil^3F(\ptil,y)$ in (\ref{HK20}). $l=1$, $\La=10$, 
$0.1\leq y<1$, $1\leq\ptil\leq 10$ . 
}
\end{minipage} 
\end{figure}

\begin{figure}
\caption{
Behaviour of the integrand of $F^-$,(\ref{HK20}). $l=1$, $\La=100$, 
$0\leq y\leq l=1$, $1\leq\ktil\leq \La=100$ . The flat plane locates at the height -0.5. 
}
\includegraphics[height=8cm]{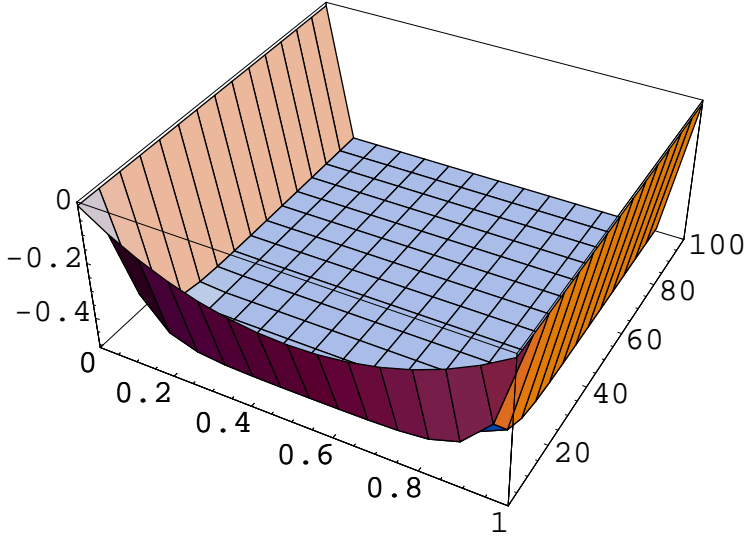}
\label{FintgrdMm1L100}
\end{figure}

The integral region, for the flat case (\ref{HK20}), is shown in 
Fig.\ref{ypINTregion}. $\m$ and $\ep$ are the IR cutoff of 
$\ptil$-axis and the UV cutoff of $y$-axis respectively. 
Fig.\ref{p3F10La} shows the integrand of (\ref{HK20}):\ 
$\ptil^3(F_f^-+4F_f^+)\equiv \ptil^3 F(\ptil,y)$. 
In Fig.\ref{FintgrdMm1L100} the behavior of the integrand of $F^-$ (\ref{HK20}) is 
shown. The (inverse) table shape says the "Rayleigh-Jeans" dominance because 
Casimir energy density is proportinal to the cubic power of $\ptil$ in the region 
$\ptil\ll \La$.                 
Numerically $E_{Cas}(\La,l)$ is obtained as 
\bea
E_{Cas}(\La,l)=\frac{2\pi^2}{(2\pi)^4}\left[ -0.1249 l\La^5
-(1.41, 0.706, 0.353)\times 10^{-5}~l\La^5\ln (l\La)\right]
\pr
\label{UIreg5}
\eea
In Fig.\ref{ypINTregion}, the region below 
the hyperbolic curve is that of Randall-Schwartz(RS)\cite{RS01}. 
They claimed the rectangular region should be replaced by this restricted one 
for the purpose of reducing the divergences. 
\bea
E^{RS}_{Cas}=
\frac{2\pi^2}{(2\pi)^4}\int_{1/l}^{\La}dq\int_{1/\La}^{1/q}dy~q^3 F(q,y)
=\frac{2\pi^2}{(2\pi)^4}[-8.93814\times 10^{-2}~\La^4]
\com
\label{surfM1}
\eea
which shows slightly milder than (\ref{UIreg5}). $\ln(l\La)$-term 
does not appear.

\begin{figure}[h]
\begin{minipage}{18pc}
\includegraphics[width=18pc]{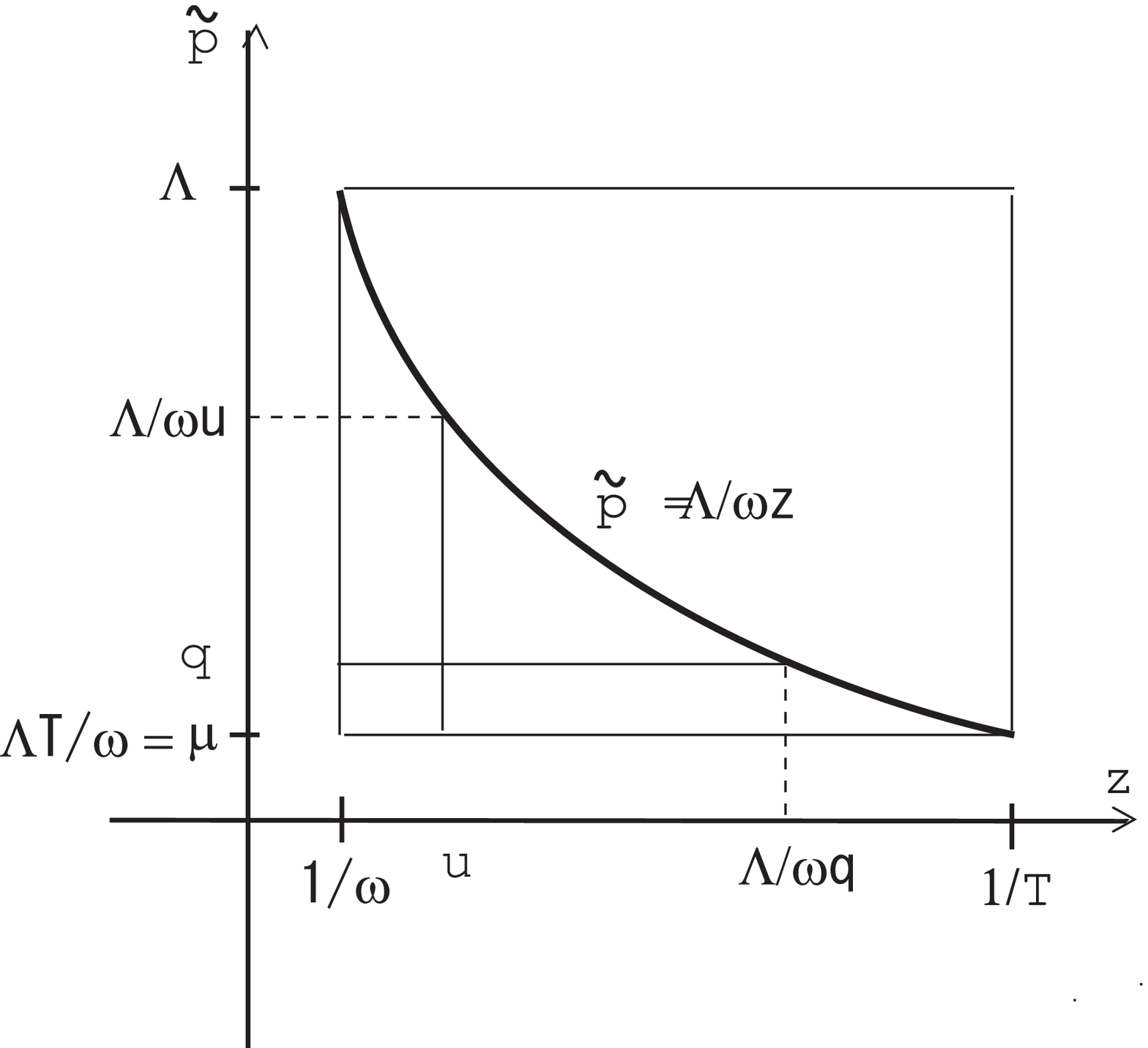}
\caption{\label{zpINTregionW}Space of (z,$\ptil$) for the integration (\ref{HKA11}). 
The hyperbolic curve 
is used in (\ref{surfM1b}).
}
\end{minipage}\hspace{2pc}%
\begin{minipage}{18pc}
\includegraphics[width=18pc]{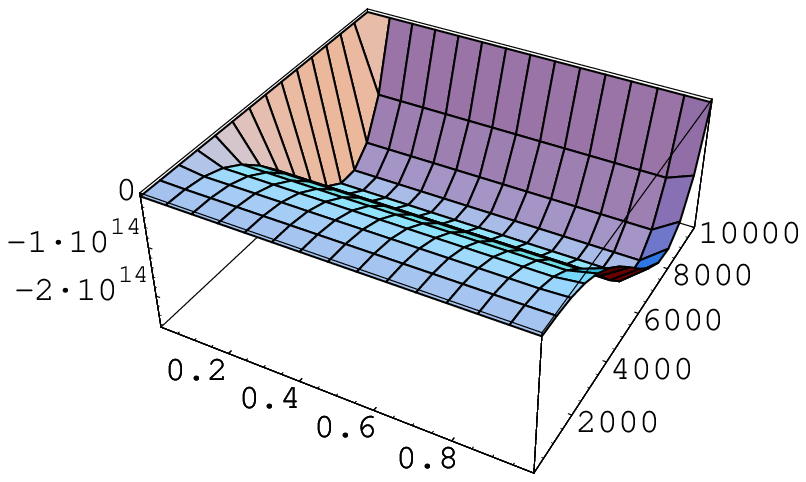}
\caption{\label{p3FmL10000}Behaviour of $(-1/2)\ptil^3F^-(\ptil,z)$ in (\ref{HKA11}). $T=1, \om=10^4, \La=10^4$.  
$1.0001/\om\leq z<0.9999/T$, $\La T/\om\leq\ptil\leq \La$.
}
\end{minipage} 
\end{figure}

\begin{figure}
\caption{
Behavior of $\ln |\half\Fcal^-(\ktil,z)|=\ln |\ktil~ G^-_k(z,z)/(\om z)^3|$. 
$\om=10^4, T=1, \La=2\times 10^4$. $1.0001/\om \leq z \leq 0.9999/T$. $\La T/\om \leq \ktil \leq \La$. 
Note $\ln |(1/2)\times (1/2)|\approx -1.39$.
}
\includegraphics[height=8cm]{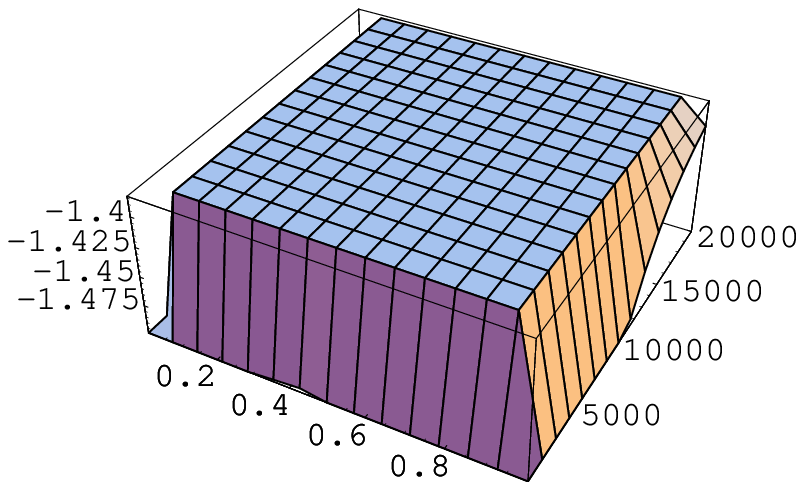}
\label{FcalmHT1k10p4}
\end{figure}

As for the Warped model, the situation is similar. Fig.\ref{zpINTregionW} shows 
the integral region. The z-axis range is $\om^{-1}\leq z\leq T^{-1}$. 
We take IR-regularization-point of $\ptil$ as $\mu=\La T/\om$. 
In Fig.\ref{p3FmL10000} the behavior of the integrand of $E^-_{Cas}$ (\ref{HKA11}) 
is shown. Fig.\ref{FcalmHT1k10p4} shows the warped version of Fig.\ref{FintgrdMm1L100}. 
$E^-_{Cas}$ (\ref{HKA11}) is numerically obtained as 
\bea
E^{\La,-}_{Cas}(\om,T)=\frac{2\pi^2}{(2\pi)^4}\times\left[ -0.0250 \frac{\La^5}{T} \right] 
\com\label{UIreg5x}
\eea
which does not depend on $\om$ and has no $\ln(\La/T)$ term. When restricted 
to RS-region, the above value changes to 
\bea
E^{-RS}_{Cas}(\om,T)=
\frac{2\pi^2}{(2\pi)^4}\int_{\m}^{\La}dq\int_{1/\om}^{\La /\om q}dz~q^3 F^- (q,z)\nn
=\frac{2\pi^2}{(2\pi)^4}\frac{\La^5}{\om}\left\{
-1.58\times 10^{-2}-1.69\times 10^{-4}\ln~\frac{\La}{\om}
                                          \right\}
\com
\label{surfM1b}
\eea
which is independent of T and has the $\ln(\La/\om)$ term. Divergence behavior 
is the same as the unrestricted case (\ref{UIreg5x}). 

\section{Sphere Lattice Regularization}
\begin{figure}[h]
\begin{minipage}{18pc}
\includegraphics[width=18pc]{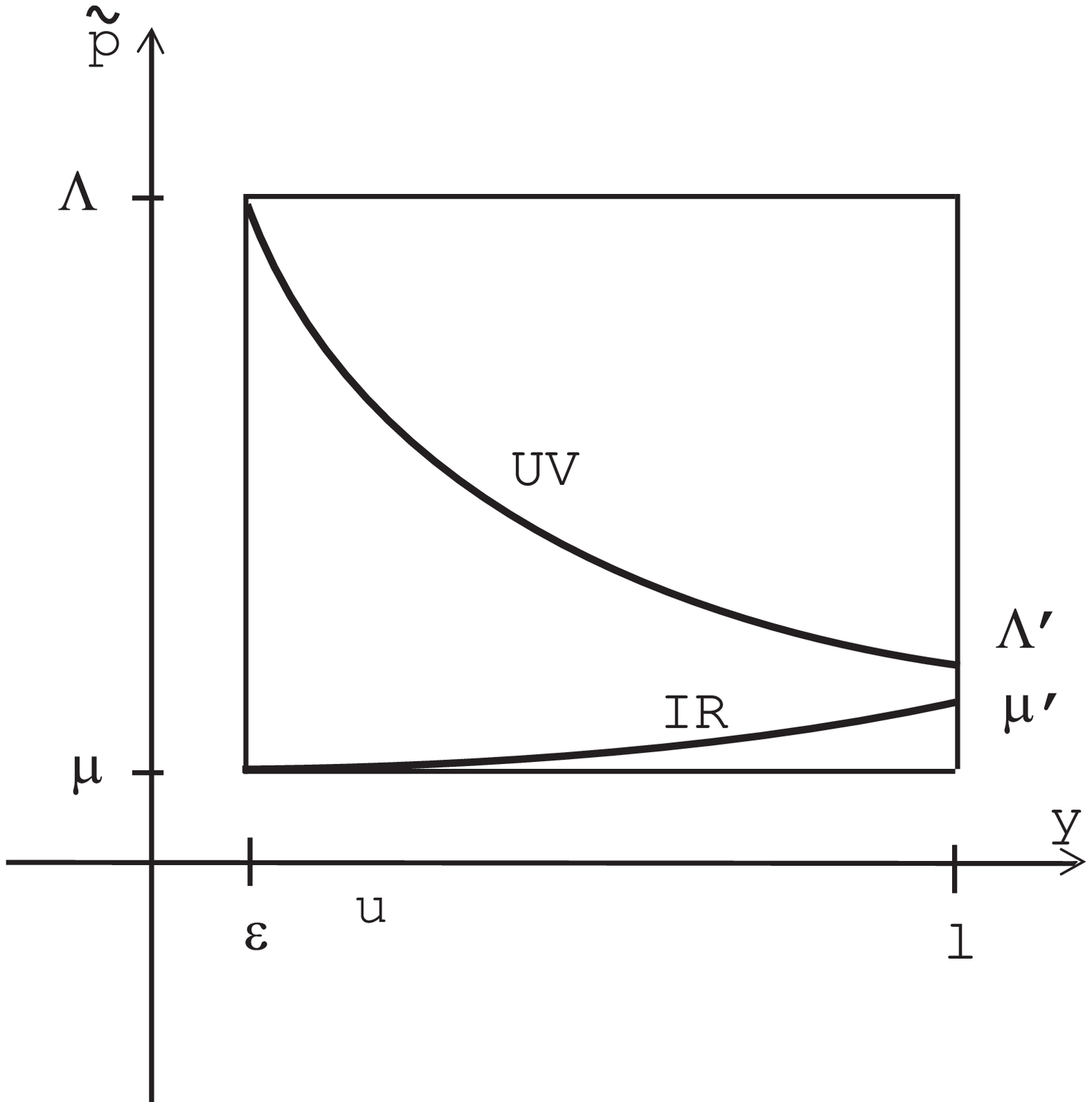}
\caption{\label{ypINTregion2}
Space of ($\ptil$,y) for the integration (present proposal as the substitute of 
Fig.\ref{ypINTregion}).
}
\end{minipage}\hspace{2pc}%
\begin{minipage}{18pc}
\includegraphics[width=18pc]{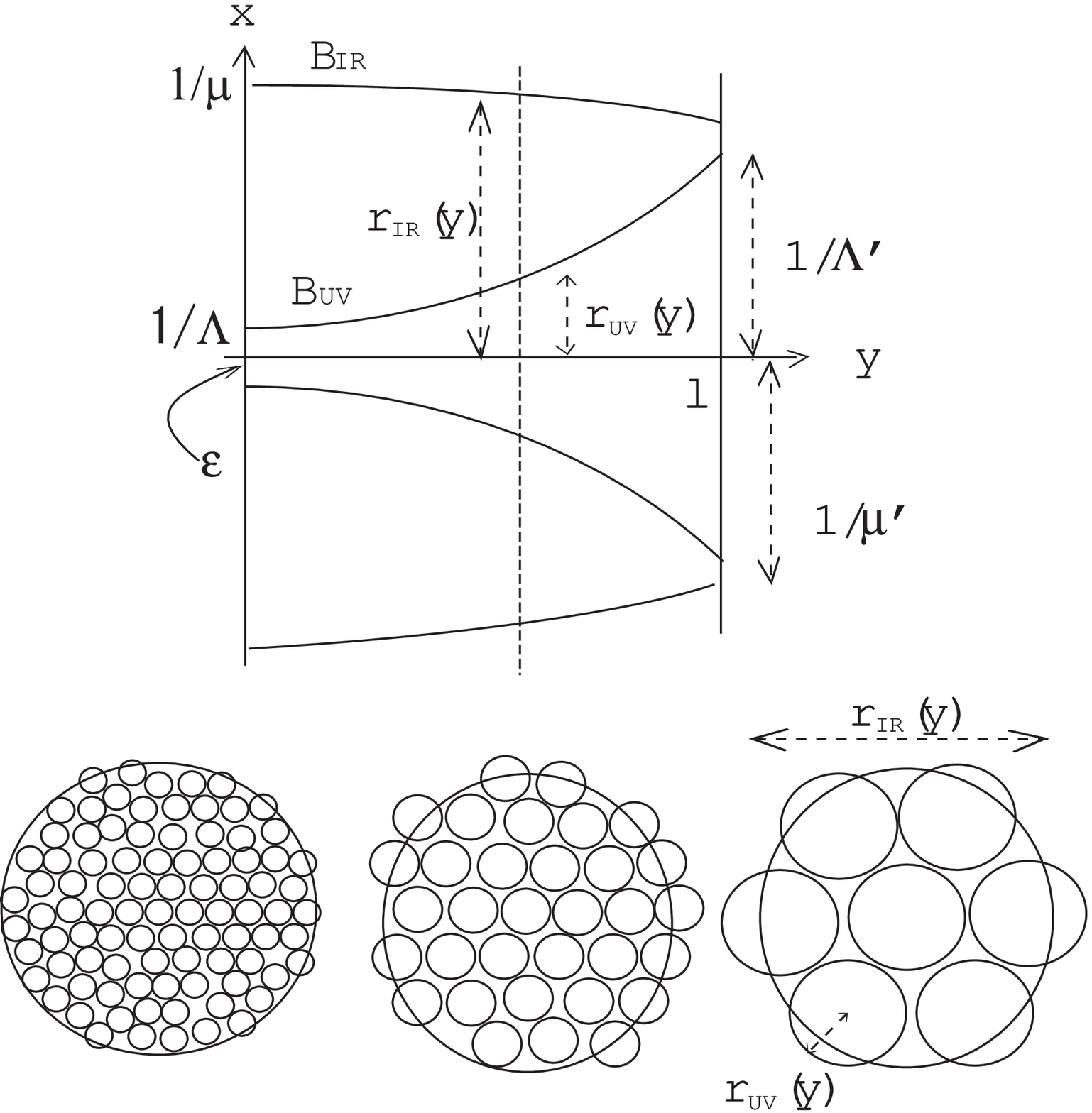}
\caption{\label{IRUVRegSurf}
Regularization Surface $B_{IR}$ and $B_{UV}$ in the 5D coordinate space $(x^\m,y)$, 
Flow of Coarse Graining (Renormalization) and Sphere Lattice Regularization. 
}
\end{minipage} 
\end{figure}

\begin{figure}[h]
\begin{minipage}{18pc}
\includegraphics[width=18pc]{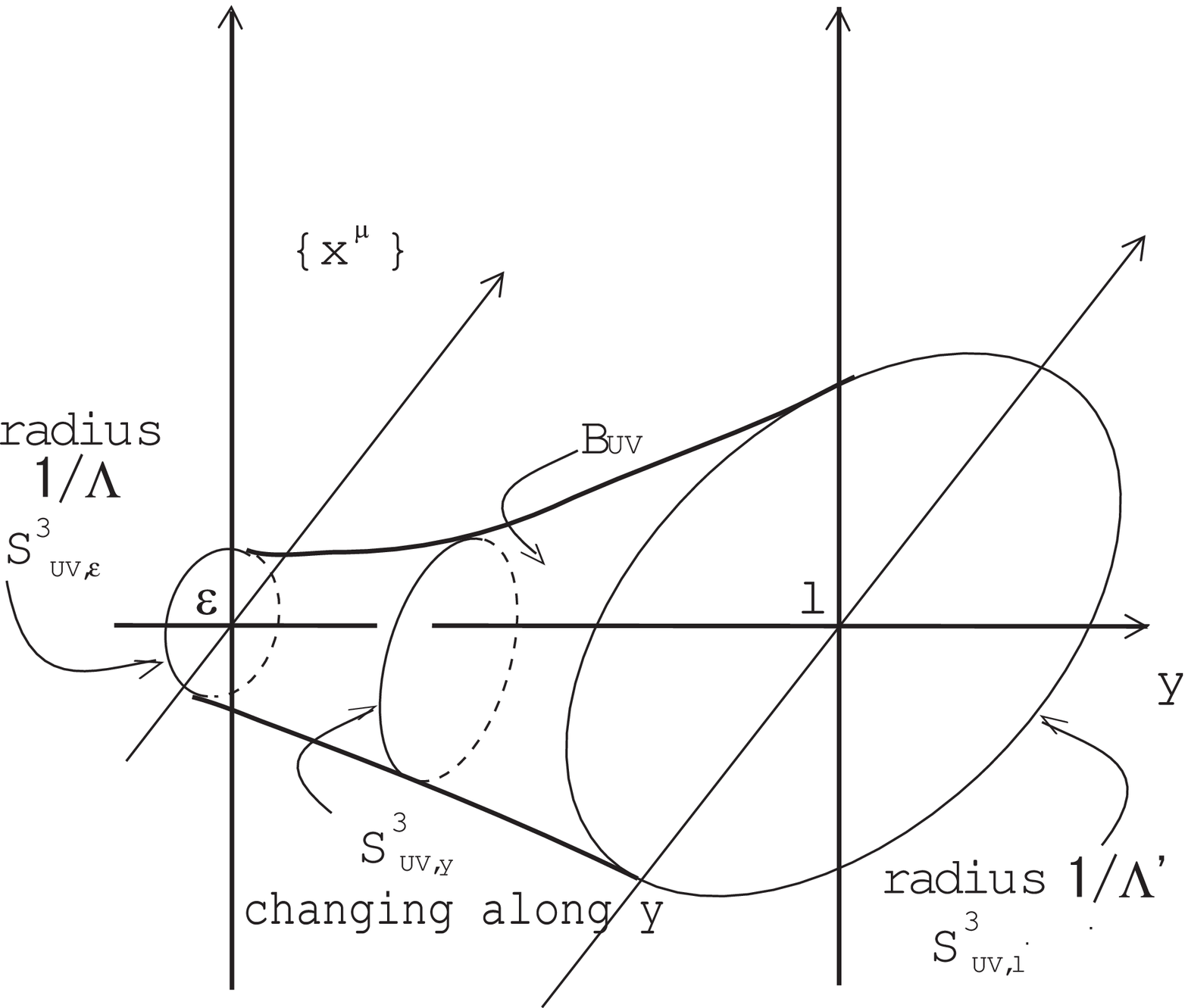}
\caption{\label{UVsurface}
UV regularization surface in 5D coordinate space.
}
\end{minipage}\hspace{2pc}%
\begin{minipage}{18pc}
\includegraphics[width=18pc]{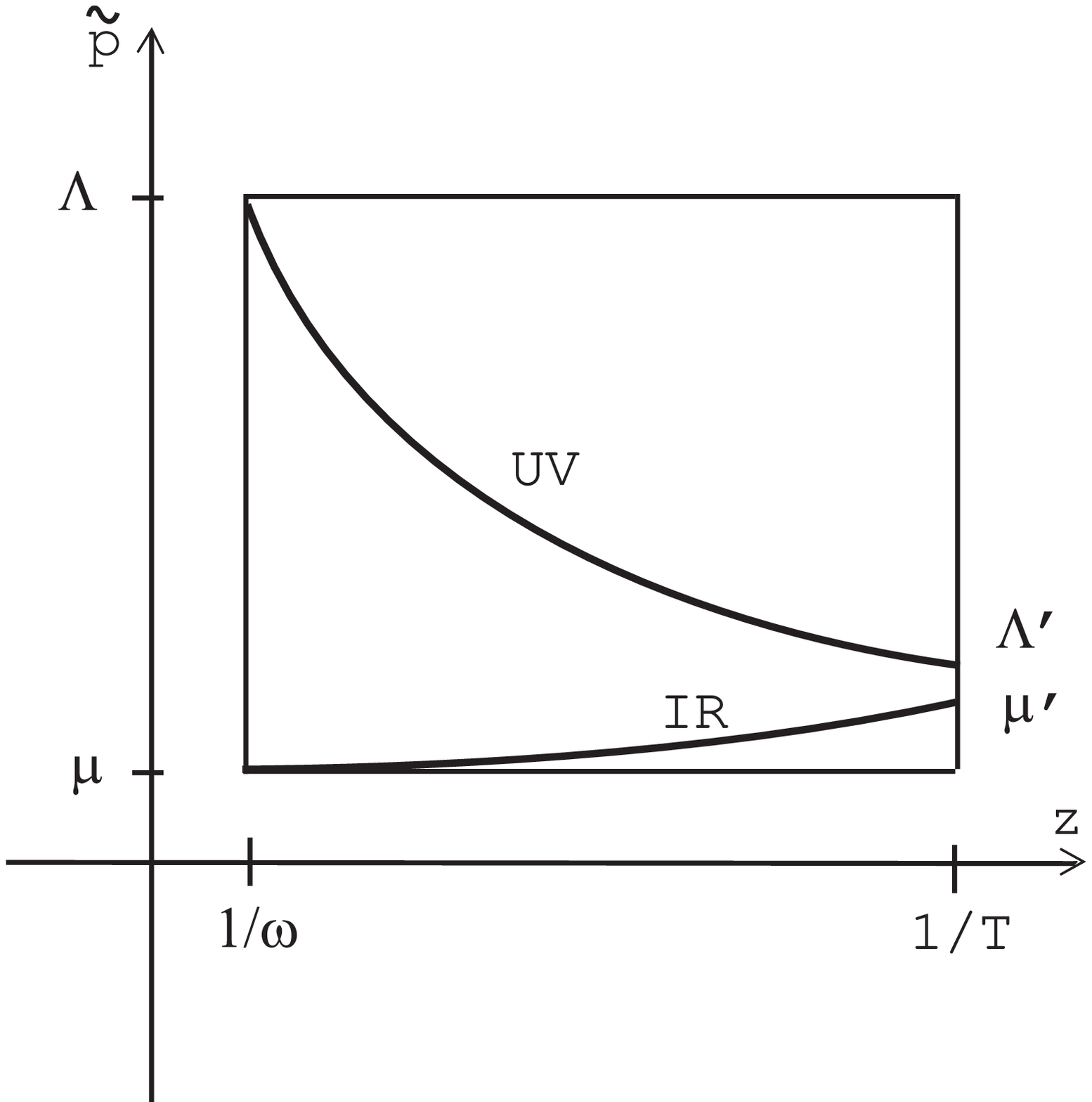}
\caption{\label{zpINTregionW2}
Space of ($\ptil$,z) for the integration (present proposal as 
the substitute of Fig.\ref{zpINTregionW}). 
}
\end{minipage} 
\end{figure}

The Randall-Schwartz's way of restricting the integral region does not sufficiently 
work to reduce the divergences. In Ref.\cite{IM0703}, we proposed 
a new way of the restriction based on the isotropy property of the system 
and the minimal area principle. Let us introduce 
two 4D hyper-surfaces, 
B$_{UV}$ and B$_{IR}$, in the 5D bulk space-time. 
\bea
\mbox{B}_{UV}\q:\q\ptil^{-1}=\sqrt{(x^1)^2+(x^2)^2+(x^3)^2+(x^4)^2}=r_{UV}(y)\com
\q \ep=\frac{1}{\La}<y<l\com\nn
\mbox{B}_{IR}\q:\q\ptil^{-1}=\sqrt{(x^1)^2+(x^2)^2+(x^3)^2+(x^4)^2}=r_{IR}(y)
\com\q \ep=\frac{1}{\La}<y<l\com
\label{surf1}
\eea
where $x^a\ (a=1,2,3,4)$ is Euclidian coordinate. 
See Fig.\ref{ypINTregion2} and Fig.\ref{IRUVRegSurf}. The surfaces, in 
the "brane" located at $y$, are 3D sphere with the radius 
$r(y)$. The radius changes along the extra axis $y$. 
The function form of $r(y)$ is given by the minimal area (of the hyper surface) 
principle. 
We integrate the region bounded 
by  B$_{IR}$ from below and by B$_{UV}$ from above as shown 
in Fig.\ref{ypINTregion2}. Some regularization parameters 
($\La,\mu,\La',\mu',\vep$) 
are defined there. The renormalization group interpretation 
is shown in Fig.\ref{IRUVRegSurf}. The surface B$_{UV}$ is 
stereo-graphically shown in Fig.\ref{UVsurface}. It shows 
the present regularization utilizes the {\it closed-string} configuration. 
The integral region for the warped case is shown in Fig.\ref{zpINTregionW2}. 

\begin{figure}
\caption{
Behaviour of $\ptil^3W_1(\ptil,y)F(\ptil,y)$(elliptic suppression). 
$\La=10,\ l=1$\ . 
$1/\La\leq y\leq 0.99999 l,\ 1/l\leq \ptil\leq \La$. 
}
\includegraphics[height=8cm]{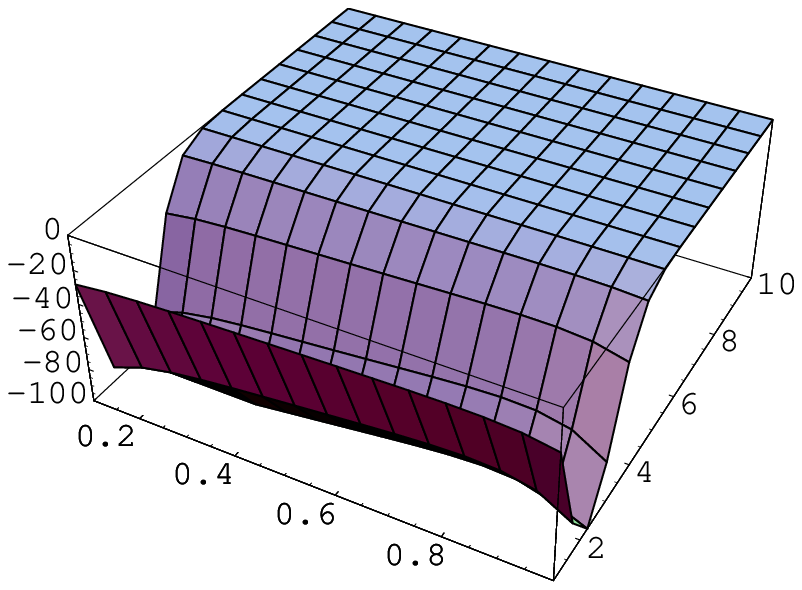}
\label{GraphW1L10m1}
\end{figure}

\section{Introduction of the Weight Function $W(\ptil,y)$ or $W(\ptil,z)$}
Another way to reduce the divergences is, instead of restriction, to introduce 
the {\it weight function} $W(\ptil,y)$ or $W(\ptil,z)$ in the integration 
over 5D space-time.  The determination of the function form is explained 
later using the minimal area principle. Now we assume some forms as sample 
examples, and check whether Casimir energy is finitely obtained. 
For the flat case, we take the following forms. 
\bea
W(\ptil,y)=  \mbox{\hspace{11cm}}\nn
\left\{
\begin{array}{cc}
(N_1)^{-1}\e^{-(1/2) l^2\ptil^2-(1/2) y^2/l^2}\equiv W_1(\ptil,y),\ N_1=1.557/8\pi^2 & \mbox{elliptic suppression}\\
(N_2)^{-1}\e^{-\ptil y}\equiv W_2(\ptil,y),\ N_2=2(l\La)^3/8\pi^2                   & \mbox{hyperbolic suppression1}
\end{array}
           \right.
\label{uncert1}
\eea
$W_2$ is considered to imitate the Randall-Schwartz restriction. 
Fig.\ref{GraphW1L10m1} shows $W_1$-weighted case of Fig.\ref{p3F10La}. 
We notice, in the "valley", the depth, the location and the shape change. 
The Casimir energy is numerically obtained as 
\bea
E^W_{Cas}=\mbox{\hspace{10cm}}\nn
\left\{
\begin{array}{cc}
-(2.500,2.501,2.501)\frac{\La}{l^3}+(-0.142,1.09,1.13)\times 10^{-4}\frac{\La\ln (l\La)}{l^3} & \mbox{for}\q W_1 \\
-(6.0392,6.0394,6.03945)\times 10^{-2} \frac{\La}{l^3}-(24.7,2.79,1.60)\times 10^{-8}\frac{\La\ln (l\La)}{l^3}&\mbox{for}\q W_2
\end{array}
           \right.
\label{uncert1b}
\eea
Triplet data show unstableness of the value due to insufficient range of 
$l$ and $\La$. $\ln(l\La)$ term vanishes for $W_2$ in the present calculational 
precision. 

\begin{figure}
\caption{
Behavior of $(-N_1/2)\ptil^3W_1(\ptil,z)F^-(\ptil,z)$(elliptic suppression). 
$\La=20000,\ \om=5000,\ T=1$\ .  
$1.0001/\om\leq z\leq 0.9999/T ,\ \m=\La T/\om\leq \ptil\leq \La$.
}
\includegraphics[height=8cm]{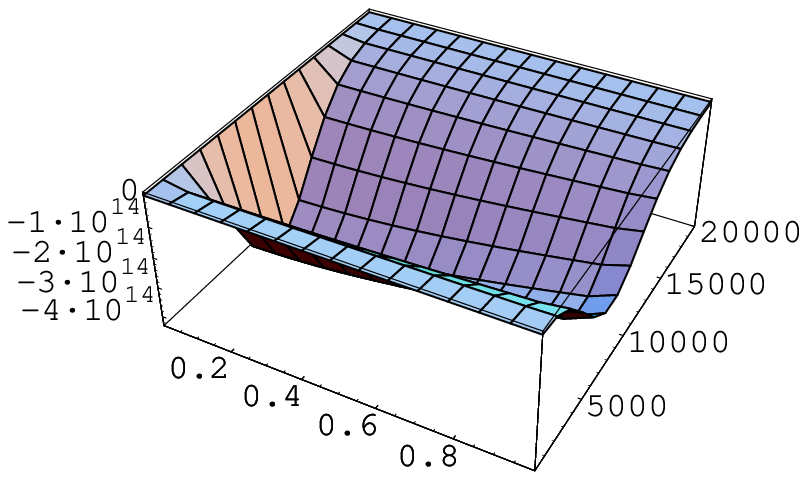}
\label{W1L2mank5senT1}
\end{figure}

For the warped case. the situation goes similarly. We take the following 
weights. 
\bea
W(\ptil,z)=\hspace{5cm}\nn
\left\{
\begin{array}{cc}
(N_1)^{-1}\e^{-(1/2) \ptil^2/\om^2-(1/2) z^2 T^2}\equiv W_1(\ptil,z),\ N_1=1.711/8\pi^2 & \mbox{elliptic suppr.}\\
(N_{2})^{-1}\e^{-\ptil zT/\om}\equiv W_2(\ptil,z),\ N_2=2\frac{\om^3}{T^3}/8\pi^2                   & \mbox{hyperbolic suppr.1}
\end{array}
           \right.
\label{uncert1x}
\eea
Fig.\ref{W1L2mank5senT1} is W$_1$-weighted case of Fig.\ref{p3FmL10000}.
The "valley" changes in its depth, shape and location. 
Casimir energy is given by 
\bea
-E^W_{Cas}=
\left\{
\begin{array}{cc}
\frac{\om^4}{T}\La\times 1.2\left\{  1+0.11~\ln\frac{\La}{\om}-0.10~ \ln\frac{\La}{T}  \right\} & \mbox{for}\q W_1 \\
\frac{T^2}{\om^2}\La^4\times 0.062\left\{  1+0.03~\ln\frac{\La}{\om}-0.08~ \ln\frac{\La}{T}  \right\}   &\mbox{for}\q W_2
\end{array}
           \right.
\label{uncert1bX}
\eea
For the hyperbolic case ($W_2$), the divergence is slightly milder than the case of RS-restriction 
(\ref{surfM1b}) ($\La^5\ra\La^4$).
It is new that $\ln(\La/T)$ appears. 

After calculating for 13 different weights\cite{SI0801, SI0812}, we conclude 
Casimir energy can be expressed as follows (except the hyperbolic weights).  
For the flat case, 
\bea
E^W_{Cas}/\La l =-\frac{\al}{l^4}\left( 1-4c\ln (l\La) \right) \pr
\label{uncert1cx}
\eea
For the warped case
\bea
E^W_{Cas}/\La T^{-1} =-\al \om^4\left( 1-4c\ln (\La/\om) -4c'\ln (\La/T) \right) 
\pr
\label{uncert1c}
\eea
The parameters $\al, c, c'$ depend on the form of the chosen weight. The factors 
$\La l$ and $\La T^{-1}$ are the area of the rectangular regions (normalization 
constants). 
This result says the divergences of the 5D Casimir energy reduces to the 
log-divergence if we take into accout the weight properly. The final log-divergence 
can be renormalized into the boundary parameters $l$ (or $T$) and $\om$. 
See Sec.\ref{conc}. 

\section{Meaning of Introduction of the Weight Function (1):\ Minimal Area Principle }
In Ref.\cite{IM0703}, we presented the following way to determine the form of $W$. 
Let us explain it in the warped case. $W(\ptil,z)$ appears as
\bea
-E^{W}_{Cas}(\om,T)=\intpE\int_{1/\om}^{1/T}dz~ W(\ptil,z)F^\mp (\ptil,z)  \nn
=\frac{2\pi^2}{(2\pi)^4}\int d\ptil\int_{1/\om}^{1/T}dz~\ptil^3 W(\ptil,z)F^\mp (\ptil,z) 
\pr
\label{weight1}
\eea
The integral region is the rectangular of Fig.\ref{zpINTregionW2}. 
We express the above expression by the following path-integral. 
\bea
-E^{W}_{Cas}(\om,T)=\int\Dcal\ptil(z)\int_{1/\om}^{1/T}dz~S[\ptil(z),z]\ ,  \nn 
S[\ptil(z),z]=\frac{2\pi^2}{(2\pi)^4}\ptil(z)^3 W(\ptil(z),z)F^\mp (\ptil(z),z).
\label{weight1a}
\eea
All possible paths $\{\ptil(z)|\om^{-1}<z<T^{-1}\}$ are summed. The dominant 
contribution, in the above path-integral, is given by $\del S=0$. 
\bea
\mbox{Dominant Path }\ptil_W(z)\ :\ \q
\frac{d\ptil}{dz}=\frac{-\frac{\pl\ln(WF)}{\pl z}}{\frac{3}{\ptil}+\frac{\pl\ln (WF)}{\pl\ptil}}
\pr
\label{weight1b}
\eea
This result says the dominant path $\ptil_W(z)$ is determined by $W(\ptil,z)$. 
A concrete example is the valley bottom line in Fig.\ref{W1L2mank5senT1}.

On the other hand, there exists another path which is determined 
independently of the above dominant path.
That is the minimal area curve $r_g(z)$ which is given by
\bea
\mbox{Minimal Surface Curve }r_g(z)\ :\q
3+\frac{4}{z}r'r-\frac{r''r}{{r'}^2+1}=0\com\q
\frac{1}{\om}\leq z\leq \frac{1}{T}
\com\label{weight2}
\eea 
where $r'=\frac{dr}{dz}, r''=\frac{d^2r}{dz^2}$. 
This differential equation is obtained by 
minimizing the area $A$ of the hyper-surface (\ref{surf1}):\ $\del A=0$ 
where $A$ is given by 
\bea
ds^2=(\del_{ab}+\frac{x^ax^b}{(rr')^2} )
\frac{dx^a dx^b}{\om^2 z^2}\equiv g_{ab}(x)dx^adx^b ,\nn
A=\int\sqrt{\det g_{ab}}~d^4x
=\int_{1/\om}^{1/T}\frac{1}{\om^4z^4}\sqrt{{r'}^2+1}~r^3 dz.
\label{weight3}
\eea 
Hence this path $r_g(z)$ is determined by the {\it induced geometry} 
$g_{ab}(x)$. We require \cite{SI0801} the following relation in order 
to define $W(\ptil,z)$.  
\bea
\ptil_W(z)=\ptil_g(z)\ (=\frac{1}{r_g(z)})
\pr\label{weight4}
\eea 

In the above procedure, we have defined the integral measure $d^4p_EdzW(\ptil,z)$ 
by the 5D bulk geometry or by the 4D induced geometry.

\section{Meaning of Introduction of the Weight Function (2):\ Fluctuation of Space-Time}
\label{fluctuate} 
For the purpose of naturally introducing the idea of the previous sections, 
we newly define Casimir energy in the higher dimension 
by the following {\it generalized path-integral}. (warped case).  
\bea
-\Ecal_{Cas}(\om,T,\La)\equiv 
\int_{1/\La}^{1/\m}d\rho\int_{
\begin{array}{l}
r(1/\om)\\
=r(1/T)\\
=\rho
\end{array}                  }
\prod_{a,z}\Dcal x^a(z)\times   \nn
F(\frac{1}{r},z)
~\exp\left[ 
-\frac{1}{2\al'}\int_{1/\om}^{1/T}\frac{1}{\om^4z^4}\sqrt{{r'}^2+1}~r^3 dz
    \right],
\label{weight5}
\eea 
where $\m=\La T/\om$ is the IR-cutoff parameter, and we take 
the limit $\La T^{-1}\ra \infty$ at the final stage. 
$1/2\al'$ is the string (surface) tension parameter 
($\al'$ has the dimension of [Length]$^4$). 
$F(\ptil,z)$ comes from the quantum fluctuation of the bulk 
matter field. 

The above path-integral expression says the 4D {\it coordinates} $x^a$ 
play the role of {\it operators} of the quantum statistical mechanics
\cite{Fey72}. The extra coordinate $z$ is the parameter of 
the {\it inverse temperature}. It looks that {\it the space-time coordinates 
are fluctuating}. 

We expect the numerical or analytical evaluation of 
Casimir energy (\ref{weight5}) directly gives similar results 
explained so far.

\section{Conclusion}\label{conc}
\begin{figure}
\caption{
Graph of Planck's radiation formula.  
$ \Pcal (\be,k)=\frac{1}{(c\hbar)^3}\frac{1}{\pi^2}k^3/(\e^{\be k}-1)\ \ 
(1\leq\be\leq 2,\ 0.01\leq k\leq 10)$.
}
\includegraphics[height=8cm]{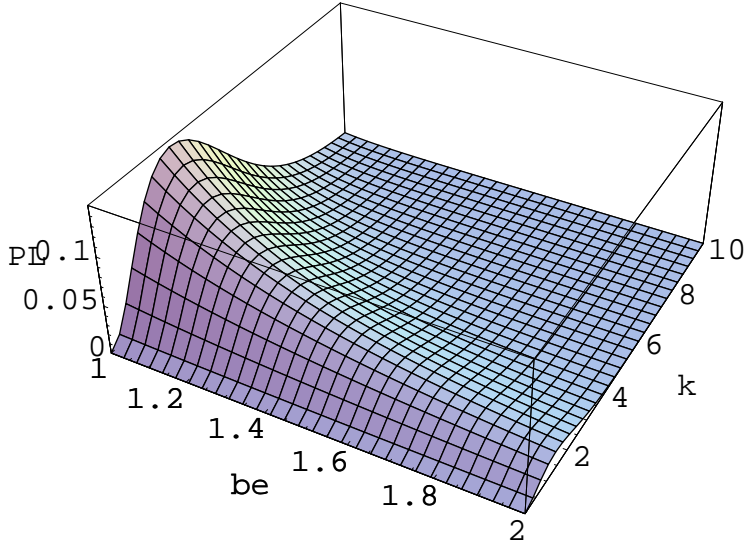}
\label{PlanckDistB}
\end{figure}

We show, in Fig.\ref{PlanckDistB}, the Planck's radiation distribution 
in the stereo-graphical way by adding the inverse temperature axis. 
The behavior looks similar to Fig.\ref{W1L2mank5senT1} except the sign. 
It says the extra coordinate corresponds to the inverse temperature. 

The log-divergences in Casimir energy (\ref{uncert1cx}) and (\ref{uncert1c}) are familiar 
in the quantum field theory. For the warped case, they are renormalized into the boundary 
parameter $\om$. 
\bea
\frac{E^W_{Cas}}{\La T^{-1}} =-\al\om^4\left( 1-4c\ln (\frac{\La}{\om})-4c'\ln (\frac{\La}{T}) \right) 
=-\al (\om_r)^4\com\nn 
\om_r=\om\sqrt[4]{1-4c\ln (\frac{\La}{\om})-4c'\ln (\frac{\La}{T}) }\pr
\label{conc1}
\eea
{\it Local counterterms are unnecessary}. Divergences are directly absorbed into 
the boundary parameter. Note that $c$ and $c'$ are pure numbers and 
represent the interaction between the bounaries and the (free) field. Compare 
this with the ordinary renormalization, like QED, where the $\be$-function 
depends on the coupling. When $c$ and $c'$ are sufficiently small, the 
$\be$-function is given as
\bea
|c|\ll 1\ ,\ |c'|\ll 1\com\q 
\om_r=\om (1-c\ln (\La/\om)-c'\ln (\La/T) )\com\nn 
\be\equiv \frac{\pl}{\pl(\ln \La)}\ln \frac{\om_r}{\om}=-c-c'
\pr
\label{conc2}
\eea
The scaling behavior of $\om$ is determined by the sign of $c+c'$. 
Because we identify Casimir energy (\ref{conc1}) with the cosmological 
constant, the sign also determines the scaling behavior of the constant. 
Note that the direction of flow, which determines the attractive or repulsive force, 
is not given by the derivative (w.r.t. the boundary parameter) of Casimir energy.  

For the flat case (\ref{uncert1cx}), the other boundary parameter $l$ 
is renormalized\cite{SI0801}. 

Parameters, which appears in the 5D warped model, can be fitted in the way 
consistent with the present observation (except the sign). 
First we take the 4D momentum cut-off $\La=M_{pl}\sim 1.2\times 10^{19}$GeV. 
The present final result (\ref{conc1}) says 
$|S(\mbox{Euclidian Action})|\sim\intfx\sqrt{-g}\la_{obs}/G_N\propto \om^4$. 
The observational data says $|S(\mbox{Euclidian Action})|\sim\intfx\sqrt{-g}\la_{obs}/G_N
\sim (R_{cos})^4 (10^{-3}\mbox{eV})^4$ where $R_{cos}\sim 5.\times 10^{41}$GeV$^{-1}$ is 
the size of the present universe. 
The experimental result about the Newton's gravitational force 
tells us the warped parameter $\om$ is taken as  
$\om\sim 10^{-3}$eV. Note that $\om\sim\sqrt{M_{pl}/R_{cos}},~R_{cos}\cdot \om\sim
\sqrt{M_{pl}\cdot R_{cos}}=\sqrt{N_{DL}}$. 
Hence $|S(\mbox{Euclidian Action})|\sim {N_{DL}}^2=
 4.\times 10^{121}$. 
From (\ref{conc1}), $|E_{Cas}|\sim \La T^{-1}\om^4$. We can 
identify $T^{-4}|E_{Cas}|\sim \La T^{-5}\om^4$ with 
$|S(\mbox{Euclidian Action})|\sim (R_{cos})^4\om^4$. Hence 
$T\sim {R_{cos}}^{-1}{N_{DL}}^{1/5}\sim 3. \times 10^{-30}$GeV. 
$\mu=\La T/\om\sim M_{pl}{N_{DL}}^{-3/10}\sim 7.$ GeV. 
The IR cutoff $\mu$ is near to the nucleon mass or the weak boson mass. 
If we interpret, in Fig.\ref{IRUVRegSurf}, the number of small 4D balls within the S$^3$ boundary 
is the degree of freedom of the present system, then it is grossly given by 
$\La^4/\mu^4=\om^4/T^4\sim {N_{DL}}^{6/5}\sim 9. \times 10^{72}$. 
Finally 
we note the neutrino mass, $m_\nu$, is similar to the warp parameter (5D bulk curvature) $\om$.

As stated in Sec.\ref{fluctuate}, the present formulation 
of the higher dimensional quantum field theory gives 
the picture of the fluctuating space-time. It is known, in the 
string theory, the uncertainty principle appears in the 
space-time coordinates\cite{Yoneya87}. 

\section{Acknowledgment}\label{ackno}
The past development is given in the proceedings of some 
conferences\cite{SI07Nara,SI0803OCU,SI0903Singa,SI0909,SI0912}. 
The author thanks the audience.

\end{document}